\documentclass[%
 reprint,
 amsmath,amssymb,
 aps,
]{revtex4-1}

\usepackage{graphicx}
\usepackage{dcolumn}
\usepackage{bm}

\begin{document}

\title{Cosmology with large galaxy redshift surveys}

\author{Laerte Sodr\'e Jr.}
\affiliation{Instituto de Astronomia, Geof\'isica e Ci\^encias 
Atmosf\'ericas \\
Universidade de S\~ao Paulo - Brazil
}
\date{\today}

\begin{abstract}
Galaxy redshift surveys are a major tool to address the most challenging
cosmological problems facing cosmology, like the nature of dark energy and
properties dark matter. The same observations are useful
for a much larger variety of scientific applications, from the study of
small bodies in the solar system, to properties of tidal streams in the
Milky Way halo, to galaxy formation and evolution. Here I briefly discuss
what is a redshift survey and how it can be used to attack astrophysical and
cosmological problems. I finish with a brief description of a new survey,
the Javalambre Physics of the Accelerating Universe Astrophysical Survey
(JPAS), which will use an innovative system of 56 filters to map $\sim8000$ square degrees on the sky.
JPAS photometric system, besides providing accurate photometric redshifts
useful for cosmological parameter estimation, will deliver a low-resolution
spectrum at each pixel on the sky, allowing for the first time an almost 
all-sky IFU science.
\end{abstract}

\maketitle


\section{Introduction}

Galaxy redshift surveys on large areas of the sky are nowadays the 
astrophysical equivalent of large high-energy physics collaborations, 
like those built around the Large Hadron Collider at CERN. They aim to build
3-D maps of the galaxy distribution, since many astrophysical and
cosmological process let their imprint in the cosmic structure traced by 
galaxies.

The investigation of the galaxy distribution obtained by
large redshift surveys is able to deliver not only reliable estimates
of cosmological parameters but, hopefully, should help to clarify the 
nature of dark energy and other cosmic problems, as more information 
is collected by new surveys, probing more deeply large areas of the sky.

Here I briefly review what are redshift surveys and what science they can
address. I start with the $\Lambda$CDM cosmology and discuss the overall
agreement between this cosmological model and the observation of galaxy
distribution. Then I discuss what are redshift surveys and present a quick
panorama of the science they plan to address.
I conclude presenting the main characteristics of the JPAS survey,
conducted by a Spanish-Brazilian collaboration and that should become
operational by 2013.

\section{Cosmology@2012}
The cosmological scene in 2012 is somewhat paradoxical. The observations
are consistent with an universe ruled by Einstein's General Relativity
with zero curvature and dominated by two mysterious components: dark energy
and cold dark matter, contributing, respectively with $\sim$3/4 and $\sim$1/4,
to the matter-energy budget required for a zero-curvature universe. The baryons
of our everyday life participate in this budget with only $\sim 4\%$, and other 
particles, like neutrinos and fotons, have a negligible contribution. 
And, although we do not
know the nature of the dominant components, we know their quantities 
with high accuracy: in terms of the density parameter the 
amount of dark energy is $\Omega_\Lambda = 0.725 \pm 0.016$ and of dark matter
is $\Omega_m = 0.274 \pm 0.012$ \citep{komatsu11}. This is the $\Lambda$CDM 
universe.The density parameter of a given component is the ratio between the 
density of this component and the critical density, $\Omega = \rho/\rho_{crit}$,
where 
the critical density is defined as $\rho_{crit} = 3H_0^2/8 \pi G$, where $H_0$ 
is the value of the Hubble constant and $G$ the gravitational constant.  

Dark energy is the name of the component responsible for the acceleration of 
the universe. It acts like an anti-gravity. 
Dark matter, on the contrary, provides
the gravitational seed which led to the formation of astronomical objects like
stars and galaxies and is, in the largest scales, the major contributor for
the gravitational pull between galaxies. 
The current cosmological paradigm assumes that these
components control the formation and evolution of large scale structures-
galaxies and groups and clusters of galaxies. Indeed, there is an excellent
overall agreement between a variety of observations and cosmological simulations
of a $\Lambda$CDM universe. Observations of the cosmic background radiation
with the satellite WMAP provide strong constraints on the curvature of the 
universe and on the amount of baryons, the latter in good agreement with
estimates obtained independently from the abundance of light elements
\citep{coc04}. Galaxy clusters provide strong evidence of dark matter, either through
gravitational lensing or X-ray emission by the hot intracluster medium
\citep[e.g.,][]{cypriano05}. It was
the study of Supernovae Ia that showed that the expansion of the universe is
accelerating, providing the first observational evidence of dark energy
\citep{riess98,perlmutter99}. 

It is this very good agreement between observations and predictions of
the $\Lambda$CDM model that imposes the need of a much deeper investigation,
aiming to understand the building bricks of our universe.

\section{The large scale structure}

In scales above several thousands of parsecs 
the universe is populated by galaxies. Galaxies 
tend to cluster, due to the attractive nature of gravitation. Our galaxy,
the Milky Way, is surrounded by many small satellites. Many bright 
galaxies are found in pairs, like the Milky Way and Andromeda in the Local 
Group; others are in more numerous groups, with a few dozen members. 
About 10\% of the galaxies are in very rich structures, the clusters of 
galaxies, with hundreds or thousands of members within a volume not much 
larger than that of the Local Group. Groups and clusters themselves  
are clustered forming much larger structures, called super-clusters, with
sizes some dozens to hundreds of times larger than the Local Group. 

The knowledge of the galaxy distribution at the largest scales is a major 
achievement of redshift surveys like the 2dF Galaxy Redshift Survey 
\citep{colless01} and 
the Sloan Digital Sky Survey \citep{abazajian05}. 
They have demonstrated that galaxies are 
distributed in a network of filaments and walls with galaxy clusters at their 
intersection. This network also contains large voids embedded, with diameters 
of a few dozens Mpc. 

But galaxies are just the visible tracers of the dominant mass component of 
the universe: cold dark matter. Cold means that the velocity of the dark 
matter particles is non-relativistic when they were formed, just after the
Big Bang. A major reason for the CDM paradigm is the strong resemblance
between the observed galaxy distribution with the large scale distribution 
of dark matter established by numerical N-body simulations \citep{springel06}. 
Indeed,
the type of dark matter has a profound effect on the appearance of the  
large scales. Cold particles allow the collapse of  very small structures, 
whereas if the universe was dominated by hot dark matter 
(e.g., massive neutrinos), only large objects, like superclusters and 
clusters, would be initially formed, and galaxies would appear later through
the fragmentation of large objects. 

The difference between the expected appearance of the universe in its
largest scales predicted by different models actually shows how powerful 
is the study of galaxy 
distribution for unveiling some of the universe mysteries.

\section{Galaxy redshift surveys}
Galaxy redshift surveys are an important tool of contemporary
astrophysics and observational cosmology. Their objective is to map the 
universe as traced by galaxies, obtaining a 3-D map of the galaxy distribution.
The reason is that the cosmological parameters are imprinted in these 3-D maps,
and their analysis is considered the most powerful way to address the nature 
of dark energy and other problems \citep{albrecht06}.

To make these 3-D maps we need to determine the cosmological distances of 
galaxies. This is usually done by measuring their spectral deviation
$z = (\lambda - \lambda_e)/ \lambda_e$.  
Due to the expansion of the universe, a foton emitted by a galaxy with 
wavelength $\lambda_e$ is observed with wavelength $\lambda$. Since in
general $\lambda > \lambda_e$ (except for some very close galaxies), the
spectral deviation is usually known as the redshift. The redshift
$z$ is directly related to the expansion: $1+z=a^{-1}$, where
$a(z)$ is the expansion factor (a growing function of $z$ with $a(0)=1$).
For a given cosmological model, the distance (there are several types of 
distances in cosmology) is a function of $z$, only.

The overall, uniform, expansion of the universe is known as the Hubble
flow, which is locally described by the Hubble law: $v_H \simeq cz = H_0 d$,
where $v_H$ is the recession velocity of a  galaxy (interpreting the spectral
deviation as due to the Doppler effect), $c$ is the velocity of light, and $d$ 
is the galaxy distance. 
But $z$ is also affected by the peculiar velocities, the movements produced
by the gravitational interaction between a galaxy and its neighbors:
$v=v_H+v_p$. In rich clusters, for example, these velocities can achieve more
than a thousand km/s. Peculiar velocities can change the value of $z$ from the
value expected for a pure Hubble flow, affecting the distance estimation
of nearby galaxies. It is worth mentioning that galaxies are a biased tracer
of dark matter, and studies of the peculiar velocity field are a powerful way to
measure this bias.

There are two ways to obtain $z$: spectroscopy and photometry. In the first
case, the measured spectrum allows the measurement of the spectral deviation  
of a galaxy with high accuracy (error in $\sigma_z \sim 3 \times 10^{-4}$). 
In the case of 
photometric redshifts, the flux in a few filters (5 for SDSS) are used as proxies for the 
spectra, and typical errors are much larger 
\citep[(e.g., $\sigma_z \sim 0.02(1+z)$ for SDSS/DR7;][]{omill12}.
Despite the large errors of photometric redshifts when compared with
spectroscopic redshifts, photometric surveys are often the choice because
photometry is much cheaper than spectroscopy in terms of telescope time
and consequently they can go wider and deeper than spectroscopic surveys.
The strength of a photometric redshift survey 
like JPAS is that the large number of narrow filters covering the optical
spectrum  increases the accuracy of estimated  photometric redshifts
by a factor of ten when compared with SDSS results.

Figure 1 \citep{benitez09} presents simulations illustrating the effect
of the peculiar velocities and redshift measurements on the derived
galaxy distribution. It also illustrates how difficult is to detect features
of the galaxy distribution in the presence of large photometric 
redshift errors. Indeed, powerful statical methods are an essential 
tool for extracting useful information from photometric observations of the 
galaxy web. An example is the Bayesian approach to estimation of photometric
redshifts \citep{benitez00} which, by the use of prior probabilities and
marginalization procedures, allows the inclusion of information often
ignored by other methods, producing results of high reliability.

\begin{figure*}
\includegraphics{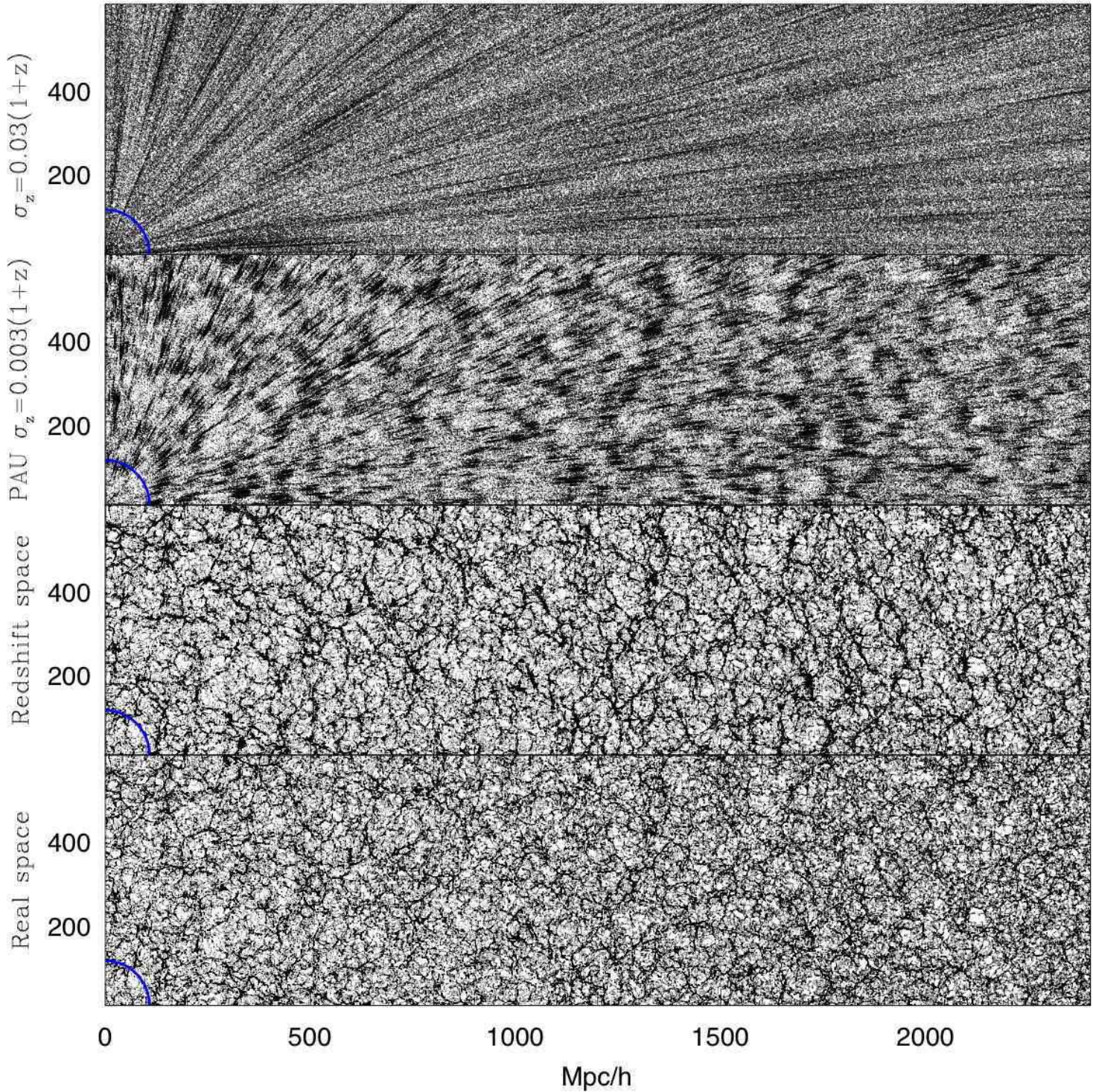}
\caption{\label{fig:wide}A simulation illustrating the effects of peculiar velocities and
photometric redshift errors \citep[from ][]{benitez09}. From bottom to top:
simulation of a galaxy distribution 1 h$^{-1}$ Mpc thick section in co-moving
coordinates a) in real space; b) in redshift space; c) in photometric
redshift space with Gaussian redshift errors of 0.003(1+z), similar to the
JPAS expectations; d) in photometric redshift space with Gaussian redshift 
errors ten times larger, 0.03(1+z). The BAO scale is shown by a section of 
circle with radius 100 h$^{-1}$ Mpc around the observer.}
\end{figure*}

\section{Cosmology with large scale probes}
The prime objective of most galaxy redshift surveys is to investigate the
nature of dark energy, mainly by constraining its equation of state,
$p_{DE}=w\rho_{DE}c^2$. The value $w=-1$ corresponds to a cosmological constant, 
which provides a good fit to the current data \citep{komatsu11}. 
But is this indeed a good model for
the behavior of dark energy or does its properties vary with redshift? 

To take in to account the possibility that dark energy evolves, 
a popular parametrization of $w$ contains a
linear expansion in the scale factor of the universe $a$: 
$w=w_0+w_a(1-a)=w_0+w_a \times z /(1+z)$ \citep[e.g.,][]{chevalier01},
adopted by the Dark Energy Task Force \citep{albrecht06} in their 
investigation of the best observational strategies to study dark energy.

In galaxy redshift surveys there are several probes to address dark energy 
properties, either based on 
distance-redshift relations or on the growth rate of cosmic structures. Indeed,
the existence of dark energy was established though the luminosity 
distance-redshift relation of type Ia supernovae \citep{riess98,perlmutter99}. 
Other canonical probes
include the baryon acoustic oscillations, cosmic shear,
and the abundance of galaxy clusters. These probes when combined with 
information obtained from the analysis of the temperature fluctuations 
of the cosmic microwave background (CMB) are able to impose stringent 
constraints on cosmological parameters \citep{eisenstein99}.

Baryon acoustic oscillations (BAOs) are acoustic waves produced during
the radiative era by the interaction of the foton-baryon plasma with dark 
matter. These waves stop propagating at recombination, when the rate of 
Compton scattering between electrons and photons becomes too low, 
and have their size frozen and equal to the size of the sound horizon at 
that epoch, $l_{BAO}\sim150h^{-1}$ Mpc \citep{eisenstein98}. 
This scale can be considered a standard ruler 
and is imprinted in the galaxy distribution. 

Since BAOs are associated with 
a density enhancement, they increase the probability of finding a galaxy at
$l_{BAO}$, which can be observed as a small excess ($\sim1$\%) in the galaxy 
correlation function \citep[e.g.,][]{percival10}. 
Analysis of SDSS luminous red galaxies  
with an algorithm that takes in to account the effects of the peculiar
velocity field \citep{padmanabhan12} 
provides a 2\% accuracy in $l_{BAO}$ in the local universe,
$z=0.35$. The measurement of this scale as a function of $z$ is a powerful 
cosmological probe  and a major objective of future surveys. 

It is worth
mentioning that the BAO features can be measured either in the 
transverse or radial directions and that each of these measurements bring
different cosmological information: radial BAO is directly sensitive to
the Hubble parameter $H(z)$, whereas transverse BAO is a probe of the
distance-redshift relation. Photometric redshift surveys are more
sensitive to transverse BAOs, since photometric redshift errors tend to blur
the radial information.

In a $\Lambda CDM$ universe structures grow from small density fluctuations
due to their gravitational atraction. 
Density fluctuations are usually expressed by the
density contrast $\delta = (\rho -\rho_m)/\rho_m$, where $\rho$ is the density 
in a given point and $\rho_m$ the mean matter density at that redshift.
For cold dark matter, these fluctuations in scales well below the Hubble 
radius evolve accordingly with the equation $\ddot \delta + 2 H \dot \delta =
4 \pi G \rho_m \delta$. For small fluctuations and at high redshifts,
$\delta$ is directly proportional to the scale factor of the universe, $a(t)$.
This is the linear phase of growth of structures. The non-linear evolution
is best studied with N-body simulations.

The largest virialized (or quasi) structures formed are the galaxy clusters. 
Since they are the largest structures to enter in the non-linear phase,
it is assumed that their material composition is 
representative of the universe as a whole. The number of clusters in a
given redshift and the cluster spatial correlations are strong functions 
of the cosmological parameters, and
the cluster mass function- the number of clusters at a given redshift with
mass in a certain interval- is then a powerful cosmological probe, since it
depends directly on the element of volume and on the growth of structures. 
Massive clusters can be found in the galaxy distribution through a variety of 
techniques \citep[e.g.,][]{wen12}. An important difficulty with this approach is the estimation
of cluster masses from the photometric information (richness and/or
luminosity) available in these surveys. This can be overcame through
multiwavelength observations (e.g., X-rays) and/or self-calibration of the
mass function \citep[][]{lima07} in combination with external mass inferences. 
Nevertheless, the results obtained up to now are consistent 
and highly complementary to those obtained by other probes 
\citep[e.g.,][]{vikhlinin09}.

The density fluctuations can also be studied through gravitational lensing.
The ellipticities of background galaxies increase as their light travels
towards the observer due to the gravitational deflection by the mass
distribution along the line of sight. These optical distortions are highly
correlated and are the signature of cosmic shear. The study of the cosmic 
shear as a function of redshift is called lensing tomography and is sensitive
to cosmic expansion through both geometry and the growth rate of structures
\citep[e.g.,][]{hu02}. It has the
potential of becoming an important cosmological probe in many future large area
surveys and, in particular, in EUCLID mission.

Gravity affects both the overall expansion and the formation and 
evolution of structures, but in completely different ways, what allows to use
observations of the 3-D galaxy distribution to test the gravitational theory
in cosmological scales. Recently, the combination of cosmic shear, galaxy
clustering and structure growth rate allowed the comparison between
general relativity and modified gravity theories, suggesting that general
relativity is a better descriptor of the behavior of gravitation theory in
large scales than some versions of alternative gravity theories 
\citep{reyes10}. Indeed,
one of the classical tests of general relativity- the gravitational redshift-
was recently measured in galaxy clusters \citep{wojtak11}, 
but the large errors still
precludes its use for evaluating alternative gravitation theories.

Massive neutrinos also let their imprint on the galaxy distribution. 
They are a kind of hot dark matter, since they are relativistic when formed.
Consequently, they escape from density fluctuations 
and, since they carry mass, they dissipate small
density fluctuations. This process, called ``free streaming'', produces a
cut-off in the number density of small fluctuations that can be detected in
the power spectrum of the galaxy distribution and, in combination with other
probes, provides strong constraints on the sum of the mass of neutrinos
species. For example, analysis of the Canada-France-Hawaii Telescope
Legacy Survey Wide Fields combined with WMAP7 data and a prior on the Hubble
constant gives a very stringent upper limit on the sum of mass of neutrino 
species equal to 0.29 eV  \citep{xia12}.

Another major scientific contribution from large scale redshift surveys is
on our knowledge on how galaxies form and evolve. The reason is that the
spectra or colors collected are useful to investigate the stellar
populations and other galaxy properties \citep{cid05}. 
Thanks mainly to SDSS photometric and spectroscopic surveys, much is known
about the galaxy populations (e.g., stellar mass, luminosity, size, stellar 
populations, mean stellar ages and metallicities)
and their relation with the environment, but big uncertainties remain.
With CDM we expect that structures grow hierarchically, with merger of
structures producing larger structures. But how does it work? Why the star 
formation rate start decreasing since $z \sim 1-2$? How massive black holes 
interact with their host galaxy? In the case of  Milk Way, SDSS lead 
to the discovery of many satellite galaxies 
\citep[e.g.,][]{belokurov10} and tidal tails, 
which are snapshots of merging activity.

\section{JPAS}
The Javalambre Physics of the Accelerating Universe Astrophysical Survey
(JPAS) is a collaboration between Spain and Brazil aiming to conduct a
survey on $\sim 8000$ square degrees with  54 narrow band filters and
two broad band filters over the range $\sim 3500$\AA ~ to  $\sim 10000$\AA,
with photometric depth $I < 22.5$. This innovative filter system actually
provides a low resolution spectrum at each pixel on the sky and was designed
to produce photometric redshifts with accuracy $\sim 0.003(1+z)$, about ten
times better than that possible with SDSS. This survey should start in 2013
and its main part should be concluded by 2018.

The survey will be conducted from the Javalambre Astrophysical Observatory
(JAO), which is being built on Pico del Buitre, near the city of Teruel, in 
Spain. The Sierra de Javalambre is amongst the darkest regions in Europe and 
has an excellent seeing, with a median of 0.71 arcsec \citep{moles10}. 
The main telescope, T250, has diameter
of 2.5m and a
very large field of view, $\sim$5 square degrees. It will be equipped with
a 1.2 Gigapixel camera wich will be a mosaic of 14 9216 x 9232 e2v CCDs, with the 
56 filters mounted in 4 trays. OAJ will also have a smaller telescope, T80, with
diameter of 80cm and equipped with a 2 square degree camera for calibration of
the photometric system of the survey. Since this telescope and camera will
be ready for operation almost one year before the T250, we shall conduct from
next year on a 5000 square degrees survey with T80: JPLUS, for Javalambre
Photometric Local Universe Survey. Besides providing calibration for JPAS,
JPLUS will allow us to test the scientific, technological and management system
of our collaboration and will provide scientific data for many interesting
projects.

The JPAS expected accuracy in photometric redshifts is enough to allow measuring BAO 
features also in the radial direction, what makes this photometric survey 
very competitive, with a DETF figure of merit above 100. We plan
to measure the BAO scale above that allowed by galaxies ($z\sim 1.3$)
by using quasars \citep{abramo12}; these probes have number densities
large enough to allow measuring BAOs up to $z \sim 3-4$. 

Besides
cosmology, JPAS will provide scientific results in many other areas:
small bodies in the solar system, Galaxy archeology, galaxy evolution, 
quasars, clusters of galaxies. An absolutely unique aspect of JPAS is that
it will allow us for the first time to do an all-sky IFU 
(for integral field unit) science, since JPAS
will measure a low resolution spectrum at each pixel on the sky. This opens
immense opportunities for studies on galaxy structure and evolution. But,
besides these very competitive scientific perspectives we are pursuing, maybe
the most compelling results of JPAS are still unknown, as always happens when
new windows are open.

\acknowledgements{I acknowledge the support of FAPESP and CNPq to this work.}

\end{document}